\documentclass[twocolumn,showpacs,preprintnumbers,amsmath,amssymb]{revtex4}
\usepackage{graphicx}
\usepackage{dcolumn}
\usepackage{bm}

\begin{document}
\title{$\bm{Z'}$ boson decay in the $\mathbf{SU\bm{(3)}_{L}\bm{\otimes} U\bm{(1)}_{N}}$ electroweak model with heavy
leptons}
\author{D. Romero}
\email{davidromeroabad@gmail.com}%
\author{O. Ravinez}%
\email{opereyra@uni.edu.pe}
\affiliation{%
Instituto de F\'isica, Facultad de Ciencias.\\
Universidad Nacional de Ingenier\'ia, Lima, Per\'u.
}%
\begin{abstract}
Based on the expectation generated by the discovery of new particles by current colliders, we analyze the decay of the $Z'$ boson in the frame of one of the $\mathbf{SU\bm{(3)}_{L}\bm{\otimes} U\bm{(1)}_{N}}$ electroweak extensions  of the standard model. The main objective is calculate the 
decay rate of this exotic boson in the aforementioned model at the tree level. With this purpose we need to develop the gauge sector, where we find thirty-three interaction terms. Mentioned particle ($Z'$) has not yet been observed experimentally, but a large number of models predict its existence. This boson exhibits a variety of decay channels, but
we will concentrate on the bosonic sector, in particular in the new charged vector bosons $V^{\pm}$ and doubly charged
$U^{\pm\pm}$ as final products, because these are special features of the model. On the other hand, we would like to remark that this model does not account for the $Z'WW$ vertex although this decay channel is considered one of the main ways to detect the $Z'$ boson in the Tevatron.
\end{abstract}
\pacs{}
\keywords{Beyond Standard Model, Z' boson}
\maketitle

\section{\label{sec:level1}Introduction}
The electroweak sector of the standard model has many extensions in the literature, several of them predicts the existence of a new neutral heavy spin one particle called Z'. Direct searches for Z' boson are carried out at Tevatron Run II and LHC \cite{Te}, \cite{Ba},  
and indirect searches will be explore by the next international lineal collider (ILC).\\ 
In the present work we calculate the decay mode $Z'\rightarrow V^{+}V^{-}$ in the frame of the $SU(3)_{L}\otimes U(1)_{N}$ electroweak model \cite{Pisano}. This model has the interesting feature
that it does not contain the decay channel $Z'\rightarrow W^{+}W^{-}$ in opposition with many other extensions.\\
In section, II we describe the model representation, emphasizing the bosonic gauge lagrangian, as well as the scalar sector, necessary to give mass to particles by the Higgs mechanism.\\
In section III, we show the decay rate obtained for the $Z'\rightarrow V^{+}V^{-}$, where the vertex was taken from the bosonic lagrangian developed in reference \cite{Mio}.\\
In section IV, we discuss the phenomenology for the Z' boson and contrast the rate decay found for this particle with the results of other models.\\
In the Appendix we show the total bosonic lagrangian after the symmetry breaking.     

\section{\label{sec:level2}The Model}

We will focus primarily in the bosonic sector since we are interested in the $Z'$ decay, with vector bosons as final products. Due to the $SU(3)_{L}\otimes U(1)_{N}$ electroweak symmetry, we obtain nine gauge bosons, eight spin-one particles are asssociated with the $SU(3)$  group , and the other one with the $U(1)$ group. 
In this model the gauge bosons before the spontaneous symmetry breaking (SSB) are determined by the algebra of the group generators. Regardless of the matter representation chosen, these bosons are the same for the differents models \cite{Ochoa}. 
\subsection{\label{sec:level2}Leptonic Sector}

We use the following triplet representation for the left-handed leptons\cite{Pleitez}: 
\begin{eqnarray*}
L_{\ell}:
\left(
\begin{array}{c}
\nu_{e}\\
e^{-}\\
E^{+}
\end{array}\right)_{L}\;,
\ \ \ 
\left(
\begin{array}{c}
\nu_{\mu}\\
\mu^{-}\\
M^{+}
\end{array}\right)_{L}\;,
\ \ \ 
\left(
\begin{array}{c}
\nu_{\tau}\\
\tau^{-} \\
T^{+}
\end{array}\right)_{L}
\ \ \ \sim \left(3,0\right)
\end{eqnarray*}
and singlets for the right-handed leptons:
\begin{eqnarray*}
R_{\ell}:\:
\begin{cases}
\:e_{R}^{-}\,,\,\ \ \:\: \mu_{R}^{-}\,,\,\ \ \: \tau_{R}^{-}\,\,\ \sim \left(1,-1\right)\\
E^{+}_{R}\,,\,\ \ M^{+}_{R}\,,\,\ \ T^{+}_{R}\,\,\ \sim \left(1,+1\right)
\end{cases}
\end{eqnarray*} 
The charge assignation for each lepton family are determined by the Gell-Mann-Nishijima relationship:
\begin{eqnarray}\label{Gell}
\frac{Q}{e}=\frac{1}{2}\left(\lambda_{3}-\sqrt{3}\lambda_{8}\right)+N 
\end{eqnarray}
where $Q$ is the electric charge, $\lambda_{3}$ and $\lambda_{8}$ are the Gell-Mann diagonal matrices, and $N$ the hypercharge.\\
The difference with respect to the Standard Model arise in the incorporation on three new charge heavy leptons $E^{+}$, $M^{+}$ and $T^{+}$ \cite{Pleitez}. 
\subsection{\label{sec:level3}Bosonic Sector}

In this section we develop the bosonic sector from the $SU(3)$ local invariant Yang-Mills lagrangian type: 
\begin{eqnarray*}
{\cal{L}}_{B}=-\frac{1}{4}\:F^{\:a}_{\mu\nu}\: F^{a\mu\nu}-\frac{1}{4}B_{\mu\nu}B^{\:\mu\nu},\ \ \ \ \  a=1,...,8.
\end{eqnarray*}
The local gauge invariance requiere that the tensors takes the form:
\begin{eqnarray*}
F^{a}_{\mu\nu}&=&\partial_{\mu}A^{a}_{\nu}-\partial_{\nu}A^{a}_{\mu}+g\:f_{abc}\:A^{b}_{\mu} A^{c}_{\nu}\\ 
B_{\mu\nu}&=&\partial_{\mu}B_{\nu}-\partial_{\nu}B_{\mu}
\end{eqnarray*}
The structure constants $f_{abc}$, are determined by the Gell-Mann matrices algebra:
\begin{eqnarray*} 
f_{abc}=\frac{1}{4i}\hbox{Tr}\left\{\left[\lambda_{a}, \lambda_{b}\right] \lambda_{c}\right\}
\end{eqnarray*}
On the other hand if we define the physics gauge bosons:
\begin{eqnarray}\label{boca}
-\sqrt{2}W^{\pm}&=&A^{1}\mp iA^{2},\nonumber\\
-\sqrt{2}V^{\pm}&=&A^{4}\pm iA^{5},\nonumber\\
-\sqrt{2}U^{\pm\pm}&=&A^{6}\pm iA^{7}.
\end{eqnarray}
it let us simplify our notation for the physics tensors:
\begin{eqnarray*}
-\sqrt{2}W^{\pm}_{\mu\nu}&=&\left(F^{1}_{\mu\nu}\mp i F^{\:2}_{\mu\nu}\right),\\
-\sqrt{2}V^{\pm}_{\mu\nu}&=&\left(F^{\:4}_{\mu\nu}\pm i F^{\:5}_{\mu\nu}\right),\\
-\sqrt{2}U^{\pm \pm}_{\mu\nu}&=&\left(F^{\:6}_{\mu\nu}\pm i F^{\:7}_{\mu\nu}\right).
\end{eqnarray*}
With this new notation the bosonic lagrangian takes the form:
\begin{eqnarray}
{\cal{L}}_{B}=&-&\frac{1}{4}\{2\:W^{+}_{\mu\nu}\: W^{\mu\nu-}+ F^{\:3}_{\mu\nu}F^{\:3\mu\nu}+2\:V^{+}_{\mu\nu}\: V^{\mu\nu-}\nonumber\\
&+&2\:U^{++}_{\mu\nu}\: U^{\mu\nu--}+F^{\:8}_{\mu\nu}F^{\:8\mu\nu}\}-\frac{1}{4}B_{\mu\nu}B^{\:\mu\nu}
\end{eqnarray}
Where the explicit development of this lagrangian in function of the physics fields is given in reference \cite{Mio}, and a list of the entire interaction lagrangian is shown in the Appendix A.
\subsection{\label{sec:level4}Scalar sector}
With the purpose of generating mass to the particles we introduce three scalar triplets:
\small
\begin{equation*}\label{rep111}
\eta=\left(\begin{array}{l}
\eta^{0}\\\\
\eta^{-}_{1} \\\\
\eta^{+}_{2}
\end{array}\right)\,\sim\left(3,0\right);\ \ \ \ 
\rho=\left(\begin{array}{l}
\rho^{+}\\\\
\rho^{0} \\\\
\rho^{++}
\end{array}\right)\,\sim\left(3,1\right);\ \ \ \ 
\end{equation*}
\begin{equation}
\chi=\left(\begin{array}{l}
\chi^{-}\\\\
\chi^{--} \\\\
\chi^{0}
\end{array}\right)\,\sim\left(3,-1\right)
\end{equation}
\normalsize
where the charge assignation is governed by Eq.(\ref{Gell}). The scalar local invariant lagrangian is written as:
\begin{eqnarray}\label{laghiggs}
{\cal{L}}_{H}&=&\left(D_{\mu}\eta\right)^{\dagger}\left(D^{\:\mu}\eta\right)+\left(D_{\mu}\rho\right)^{\dagger}\left(D^{\:\mu}\rho\right)+\left(D_{\mu}\chi\right)^{\dagger}\left(D^{\:\mu}\chi\right)\nonumber\\
&+&V\left(\eta,\:\rho,\:\chi\right)
\end{eqnarray}
We choose the more general gauge invariant potential which contains the three scalar triplets \cite{Pisano}:
\begin{eqnarray}
V\left(\eta, \rho, \chi\right)&=&\mu^{2}_{1} \eta^{\dag} \eta +\mu^{2}_{2} \rho^{\dag} \rho+\mu^{2}_{3} \chi^{\dag} \chi +\alpha_{1} \left(\eta^{\dag} \eta\right)^{2}\nonumber\\
&+& \alpha_{2} \left(\rho^{\dag} \rho\right)^{2}
+\alpha_{3} \left(\chi^{\dag} \chi\right)^{2}
+\alpha_{4}\left(\eta^{\dag} \eta \right)\left(\rho^{\dag} \rho\right)\nonumber\\
&+&\alpha_{5}\left(\eta^{\dag} \eta \right)\left(\chi^{\dag} \chi\right)  +  \alpha_{6} \left(\rho^{\dag} \rho\right)\left( \chi^{\dag} \chi \right)\nonumber\\
&+& \sum_{ijk} \epsilon^{ijk} \left(f \eta_{i} \rho_{j} \chi_{k} +h.c\right)
\end{eqnarray}
The vacuum expectation value for the neutral scalar components are:
\small
\begin{eqnarray*} 
\left\langle \eta^{0} \right\rangle = \frac{1}{\sqrt{2}} \left(\begin{array}{l}
v_{\eta}\\\\
0 \\\\
0
\end{array}\right)
\,;\, 
\left\langle \rho^{0} \right\rangle = \frac{1}{\sqrt{2}}\left(\begin{array}{l}
0\\\\
v_{\rho} \\\\
0
\end{array}\right)\,;\, 
\left\langle \chi^{0} \right\rangle=\frac{1}{\sqrt{2}}\left(\begin{array}{l}
0\\\\
0 \\\\
v_{\chi}
\end{array}\right)
\end{eqnarray*}
\normalsize 
The covariant derivative must takes the form:
\begin{eqnarray}\label{derivada}
D^{\:\mu}\:\varphi_{i}=\left[\;\partial^{\mu}-\:\frac{i\:g}{2}\:\lambda_{j}\:A^{\mu}_{j}+\:ig'\:B^{\:\mu}\:N_{\varphi}\;\right]\:\varphi_{i}
\end{eqnarray}
where $\varphi_{i}=\eta,\:  \rho,\:   \chi$.\\
When we work the kinetic term fron the lagrangian given in Eq.(\ref{laghiggs}) it is observed that the covariant derivative acting over the scalar fields let us to obtain the coupling between the gauge bosons and the scalars. The expansion about the vacuum state let identify the bosons mass when we contrast the resulting term from this lagrangian with the proca lagrangian. So, for the charged bosons we obtain:
\begin{eqnarray*}
M^{2}_{W}=\frac{1}{4}\:g^{2}(v_{\eta}^{2}&+&v_{\rho}^{2}),\ \ \ \ \ \  M^{2}_{V}=\frac{1}{4}\:g^{2}(v_{\eta}^{2}+v_{\chi}^{2})\nonumber\\
M^{2}_{U}&=&\frac{1}{4}\:g^{2}(v_{\rho}^{2}+v_{\chi}^{2})\nonumber
\end{eqnarray*}
and for the neutral bosons we obtain the mass matrix:
\small
\begin{eqnarray*}
M^{2}=\frac{g^{2}}{4}\bordermatrix{&   \cr  
&  \displaystyle v_{\eta}^{2}+v_{\rho}^{2}                   &    \displaystyle\frac{1}{\sqrt{3}}(v_{\eta}^{2}-v_{\rho}^{2})   &    \displaystyle-2\frac{g'}{g}v_{\rho}^{2} \cr 
&   \displaystyle\frac{1}{\sqrt{3}}(v_{\eta}^{2}-v_{\rho}^{2})  &  \displaystyle\frac{1}{3}(v_{\eta}^{2}+v_{\rho}^{2} +4v_{\chi}^{2})            & \displaystyle\frac{2}{\sqrt{3}}\frac{g'}{g}(v_{\rho}^{2} +2v_{\chi}^{2}) \cr
&  \displaystyle-2\frac{g'}{g} v_{\rho}^{2}     & \displaystyle\frac{2}{\sqrt{3}}\frac{g'}{g}(v_{\rho}^{2}  +2v_{\chi}^{2})  & \displaystyle4\frac{g'^{2}}{g^{2}}(v_{\rho}^{2} +v_{\chi}^{2}) \cr}
\end{eqnarray*}
\normalsize
Having obtained a non-diagonal matrix for the neutral fields is a signal that these fields are not the physical fields that are expected to find in nature, then to find them, it will be necessary diagonalize the mass matrix. The eigenvectors of this matrix give us the real physical fields and the eigenvalues their masses: 
\begin{eqnarray*}
M^{2}_{A}=0,\ \ \ \  \ \ \ \ \ M^{2}_{Z'}\approx \frac{g^{2}}{3}\left(\frac{c_{W}^{2}}{1-4s_{W}^{2}}\right)v^{2}_{\chi}
\end{eqnarray*}
\begin{eqnarray}
M^{2}_{Z} \approx \frac{g^{2}}{4}\left(\frac{1}{c_{W}^{2}}\right)\left(v^{2}_{\eta}+v^{2}_{\rho}\right)
\end{eqnarray}
The relationship between the physical fields and the gauge fields is given by:
\begin{eqnarray*}
A^{3}_{\mu}&=& s_{W}\:A_{\mu}+c_{W}\:Z_{\mu}\nonumber\\
A^{8}_{\mu}&=& -\sqrt{3}\:s_{W}\:A_{\mu}- \sqrt{3}\: s_{W}\:t_{W}\:Z_{\mu}-\frac{\sqrt{1-4s_{W}^{2}}}{c_{W}}\:Z_{\mu}'\nonumber\\
B_{\mu}&=& \sqrt{1-4s_{W}^{2}}\:A_{\mu}-t_{W}\:\sqrt{1-4s_{W}^{2}}\:Z_{\mu}-\sqrt{3}\:t_{W}\:Z_{\mu}'
\end{eqnarray*}
where $t=g'/g\equiv \tan\theta$ y $s_{W}=\sin\theta_{W}=t/(1+4\:t^{2})^{1/2}$.\\
The coupling constants introduced by the model with the covariant derivative let identify the coefficient of the electromagnetic interaction as the electron charge:  
\begin{eqnarray*}
e=\frac{g \sin\theta}{(1+3\sin^{2}\theta)^{1/2}}=\frac{g' \cos\theta}{(1+3\sin^{2}\theta)^{1/2}}\nonumber=g\sin\theta_{W}
\end{eqnarray*}

\section{ $\bm{Z'}$ Boson decay at the tree level: unpolarized process}
Here we present the corresponindg Feynman's diagram for the decay  $Z'\:\rightarrow\: V^{+}+V^{-}$ at tree level:

\begin{figure}[h]
\begin{center}
\includegraphics*[width=0.40\textwidth,height=0.30\textwidth]{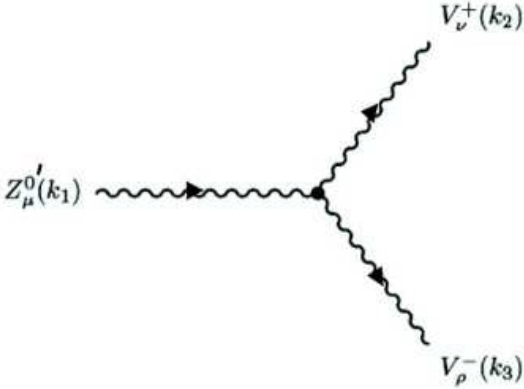}
\end{center}
\caption{\label{fig1}$Z'\:\rightarrow\: V^{+}+V^{-}$ decay.}
\end{figure}
\noindent
The amplitude of the process is determined by
\begin{eqnarray*}
\mathcal{M}=\mathcal{M}_{\mu \nu \rho}\:\epsilon^{\mu}(k_{2},\lambda_{2})
\:\epsilon^{\nu}(k_{3},\lambda_{3})
\:\epsilon^{\rho}(k_{1},\lambda_{1})
\end{eqnarray*}
and the corresponding Feynman's rule \cite{Mio}:
\begin{eqnarray*}
\mathcal{M}_{\mu\nu\rho}&=&\frac{ig}{2}\frac{\sqrt{3(1-4s^{2}_{W})}}{c_{W}}\:\times\\
&\times&\{\:g_{\mu\nu}(k_{3}-k_{2})_{\rho}
+g_{\mu\rho}(k_{2}+k_{1})_{\nu}
-g_{\nu\rho}(k_{3}+k_{1})_{\mu}\}
\end{eqnarray*}
using the approximation for high energies $M_{V}<<\left|\mathbf{k}\right| \approx M_{Z'}$ we obtain:  
\begin{eqnarray*}
\left|\mathcal{M}\right|^{2}&=& \frac{1}{3}\left|C\right|^{2}\left\{-4\left|\mathbf{k}\right|^{2}-\frac{2\left|\mathbf{k}\right|^{4}}{M_{Z'}^{2}}+\frac{5\left|\mathbf{k}\right|^{4}}{M_{V}^{2}}+\frac{16\left|\mathbf{k}\right|^{6}}{M_{V}^{4}}\right\}
\end{eqnarray*}
where
\begin{eqnarray*}
\left|C\right|^{2}=\frac{3 g^{2}}{4}\frac{\left(1-4s^{2}_{W}\right)}{c^{2}_{W}},\ \ \ \ \ \ \ \left|\mathbf{k}\right|\equiv\left|\mathbf{k}_{2}\right|=\left|\mathbf{k}_{3}\right|
\end{eqnarray*} 
Finally the rate decay mode for this process has the form:
\begin{eqnarray}\label{tasadeca11}
\Gamma\left(Z'\rightarrow V^{+}V^{-}\right)=M_{Z'}\frac{\alpha\left(1-4s_{W}^{2}\right)}{s_{2W}^{2}}\:f(x)
\end{eqnarray}
where:
\begin{eqnarray*}
f(x)=\frac{\left(\sqrt{1-x}\right)^{3}}{x^{2}}\left[\frac{1}{16}x^{3}+\frac{13}{16}x^{2}-\frac{27}{8}x+2\right],\ \ \ x \equiv \frac{4\:M^{2}_{V}}{M^{2}_{Z'}}
\end{eqnarray*}
For the case of $Z'\rightarrow U^{++}U^{--}$ the only diference resides in the mass $M_{U}$, because they have the same Feynman rule as you can see in the Appendix A, so it would have to replace $M_{V}$ by $M_{U}$ in the rate decay calculate in Eq. (\ref{tasadeca11}).
\section{Phenomenology}
The existence of the double charged bosons have in the process $\ e^{-}e^{-}\rightarrow \mu^{-}\mu^{-}$ its better experimental test \cite{Ng}.
The extension of the SM presented, is the simplest way to enlarge the gauge group without missing the natu\-ral features of the electroweak model.
From the limit $\sin^{2}\theta_{W}<1/4$ \cite{Ng}, we obtain that $M_{Z'}$ is lower than $3.1$ TeV \cite{Lang2}. The masses from the new gauge bosons $V^{\pm}$ and $U^{\pm\pm}$ are limited by the leptonic collision experiments \cite{cole} and by the muon decay \cite{mu}.
At present the lowest limit for the double charged boson is obtained from the muon-antimuon conversion $\ e^{+}\mu^{-}\rightarrow e^{-}\mu^{+} $ which give us that $M_{U^{++}}\geq 850$ GeV \cite{Will}.
By other hand the restriction $M_{V^{+}}>440$ GeV was derivated by the limits of the decay width of the muon \cite{Tully}.
According to the relationship that exist between the $M_{V}$ and $M_{Z'}$ shown in \cite{Ng}, it find the lower limit $M_{Z'}\geq 1.3$ TeV. From the mass range obtained for the  $Z'$, we can conclude that the mass for the new charged gauge bosons are less than the half of $M_{Z'}$ so the $\ Z'\rightarrow V^{+}V^{-}\ $ decay is kinematically possible  \cite{Perez},\cite{Ng}.\\
We find it interesting to compare our result obtained for the decay rate of the process $Z'\:\rightarrow\:V^{+}+V^{-}$ shown in Eq.(\ref{tasadeca11})  with another model that incorporates the same gauge symmetry. In the work of Perez, Tavares-Velasco, Toscano (PTVT) \cite{Perez}, they consider a model with $SU(3)\otimes U(1)$ gauge symmetry too, but the change is in the scalar sector, where they introduce a sextet for the Higgs fields in addition to the three triplets that we show in (\ref{rep111}), and in the fermionic sector where they use antitriplets.\\
The decay rate shown in the work of (PTVT) is:
\begin{eqnarray*}
\Gamma=\overbrace{M_{Z'}\frac{\alpha\left(1-4s_{W}^{2}\right)}{s_{2W}^{2}}}^{\kappa}\:g(x)
\end{eqnarray*}
where
\begin{eqnarray*}
g(x)=\frac{\sqrt{1-x}}{x^{2}}\left[-\frac{3}{4}x^{3}-\frac{17}{4}x^{2}+4x+1\right],\ \ x \equiv \frac{4\:M^{2}_{V}}{M^{2}_{Z'}}
\end{eqnarray*}
By comparing this expression with that found in Eq.(\ref{tasadeca11}), we observe that the difference lies in the functions $f(x)$ and $g(x)$ so it seems interesting to show deviations between them, see Fig 2:
\begin{figure}[h]
\begin{center}
\includegraphics*[width=0.50\textwidth,height=0.40\textwidth]{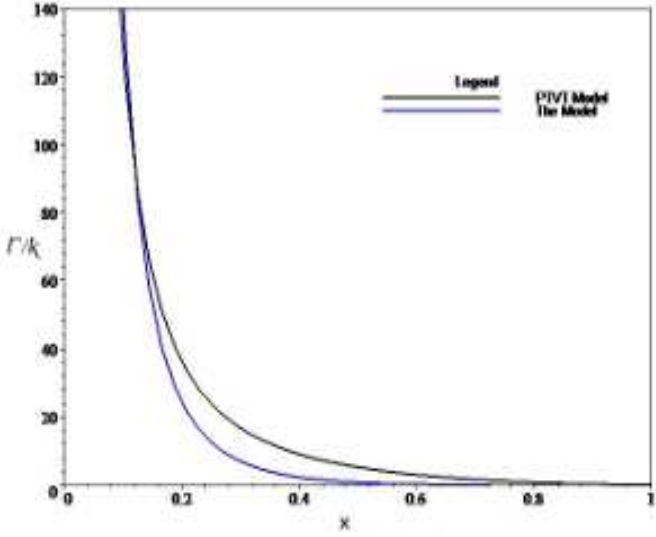}
\end{center}
\caption{\label{fig2} $Z'$ decay rate.}
\end{figure}
\section{Conclusions}
We have presented an extension of Standard Model with $SU(3)_{L}\otimes U(1)_{N}$ gauge symmetry based primarly on the Pleitez, Tonasse work \cite{Pleitez}. The resulting bosonic Lagrangian and the rate decay obtained for the $Z'$ are contrasted with the work of
Perez, Tavares-Velasco and Toscano \cite{Perez}, \cite{Tavares}. An important aspect that we highlight is the non-appearance of the interaction term between the exotic boson $Z'$ and the $W^{\pm}$ charge bosons in the bosonic lagrangian, although other models such as \cite{Perez} where this link is present. Despite that still there is no such experimental observation from this decay, this reaction is expected to be one of the possible ways to detect the $Z'$ boson in the Tevatron \cite{Naka}.\\
\begin{acknowledgments}
We would like to thank Mr. V. Pleitez for reading the manuscript and giving valuable comments, and to the Facultad de Ciencias de la Universidad de Ingenieria for the partial support.
\end{acknowledgments}
\appendix
\section{}
Below we show the total bosonic Lagrangian after the spontaneous symmetry breaking:
\begin{center}
\textbf{Trilinear vertices}
\end{center}
\small
\begin{eqnarray*}
\mathcal{L}_{\gamma WW}=ie\{(W^{\mu+}W^{\nu-}-W^{\mu-}W^{\nu+})\partial_{\mu}A_{\nu}\\
+(\partial_{\mu}W_{\nu}^{-}-\partial_{\nu}W_{\mu}^{-})W^{\nu+}A^{\mu}-(\partial_{\mu}W_{\nu}^{+}-\partial_{\nu}W_{\mu}^{+})W^{\nu-}A^{\mu}\}
\end{eqnarray*}
\begin{eqnarray*}
\mathcal{L}_{Z WW}=ig\:c_{W}\{(W^{\mu+}W^{\nu-}-W^{\mu-}W^{\nu+})\partial_{\mu}Z_{\nu}\\
+(\partial_{\mu}W_{\nu}^{-}-\partial_{\nu}W_{\mu}^{-})W^{\nu+}Z^{\mu}-(\partial_{\mu}W_{\nu}^{+}-\partial_{\nu}W_{\mu}^{+})W^{\nu-}Z^{\mu}\}
\end{eqnarray*}
\begin{eqnarray*}
\mathcal{L}_{\gamma VV}=ie\{(V^{\mu+}V^{\nu-}-V^{\mu-}V^{\nu+})\partial_{\mu}A_{\nu}\\
+(\partial_{\mu}V_{\nu}^{-}-\partial_{\nu}V_{\mu}^{-})V^{\nu+}A^{\mu}-(\partial_{\mu}V_{\nu}^{+}-\partial_{\nu}V_{\mu}^{+})V^{\nu-}A^{\mu}\}
\end{eqnarray*}
\begin{eqnarray*}
\mathcal{L}_{Z VV}=-\frac{ig}{2}\:\frac{(1+2s_{W}^{2})}{c_{W}}\{(V^{\mu+}V^{\nu-}-V^{\mu-}V^{\nu+})\partial_{\mu}Z_{\nu}\\
+(\partial_{\mu}V_{\nu}^{-}-\partial_{\nu}V_{\mu}^{-})V^{\nu+}Z^{\mu}-(\partial_{\mu}V_{\nu}^{+}-\partial_{\nu}V_{\mu}^{+})V^{\nu-}Z^{\mu}\}
\end{eqnarray*}
\begin{eqnarray*}
\mathcal{L}_{Z' VV}=\frac{ig\sqrt{3}}{2}\:\frac{\sqrt{1-4s_{W}^{2}}}{c_{W}}\{(V^{\mu+}V^{\nu-}-V^{\mu-}V^{\nu+})\partial_{\mu}Z_{\nu}'\\
+( \partial_{\mu} V^{-}_{\nu} - \partial_{\nu} V^{-}_{\mu} )V^{\nu +} Z'^{\mu}-(\partial_{\mu}V^{+}_{\nu}- \partial_{\nu}V^{+}_{\mu} )mV^{\nu-}Z'^{\mu} \}
\end{eqnarray*}
\begin{eqnarray*}
\mathcal{L}_{\gamma UU}=ie\{(U^{\mu++}U^{\nu--}-U^{\mu--}U^{\nu++})\partial_{\mu}A_{\nu}\\
+(\partial_{\mu}U_{\nu}^{--}-\partial_{\nu}U_{\mu}^{--})U^{\nu++}A^{\mu}-(\partial_{\mu}U_{\nu}^{++}- \partial_{\nu} U_{\mu}^{++})U^{\nu--}A^{\mu}\}
\end{eqnarray*}
\begin{eqnarray*}
\mathcal{L}_{Z UU}=-\frac{ig}{2}\:\frac{(1+2s_{W}^{2})}{c_{W}}\{(U^{\mu++}U^{\nu--}-U^{\mu--}U^{\nu++})\partial_{\mu}Z_{\nu}\\
+(\partial_{\mu}U_{\nu}^{--}-\partial_{\nu}U_{\mu}^{--})U^{\nu++}Z^{\mu}-(\partial_{\mu}U_{\nu}^{++}-\partial_{\nu}U_{\mu}^{++})U^{\nu--}Z^{\mu}\}
\end{eqnarray*}
\begin{eqnarray*}
\mathcal{L}_{Z' UU}=\frac{ig}{2c_{W}}\:\sqrt{3(1-4s_{W}^{2})}\{(U^{\mu++}U^{\nu--}-U^{\mu--}U^{\nu++})\partial_{\mu}Z'_{\nu}\\
+(\partial_{\mu}U_{\nu}^{--}-\partial_{\nu}U_{\mu}^{--})U^{\nu++}Z'^{\mu}-(\partial_{\mu}U_{\nu}^{++}-\partial_{\nu}U_{\mu}^{++})U^{\nu--}Z'^{\mu}\}
\end{eqnarray*}
\begin{eqnarray*}
\mathcal{L}_{WUV}=\frac{ig}{\sqrt{2}}\:\{W^{\mu+}(U^{--}_{\mu\nu}V^{\nu+}-V^{+}_{\mu\nu}U^{\nu--})-W^{+}_{\mu\nu}U^{\mu--}V^{\nu+}\\
-W^{\mu-}(U^{++}_{\mu\nu}V^{\nu-}-V^{-}_{\mu\nu}U^{\nu++})+W^{-}_{\mu\nu}U^{\mu++}V^{\nu-}\}
\end{eqnarray*}
\vspace{0.5cm}
\begin{center}
\normalsize
\textbf{Quartic vertices}
\end{center}
\vspace{0.5cm}
\footnotesize
\begin{eqnarray*}
\mathcal{L}_{W^{2} \gamma^{2} }=e^{2}\left\{W_{\mu}^{+}W_{\nu}^{-}A^{\mu}A^{\nu}-W^{+}_{\mu}W^{\mu-}A_{\nu}A^{\nu} \right\}
\end{eqnarray*}
\begin{eqnarray*}
\mathcal{L}_{W^{2} Z^{2} }=g^{2}c^{2}_{W}\left\{W_{\mu}^{+}W_{\nu}^{-}Z^{\mu}Z^{\nu}-W^{+}_{\mu}W^{\mu-}Z_{\nu}Z^{\nu} \right\}
\end{eqnarray*}
\begin{eqnarray*}
\mathcal{L}_{W^{2} \gamma Z }=e\:g\:c_{W}\{ W_{\mu}^{+}W_{\nu}^{-}(Z^{0\nu}A^{\mu}+A^{\nu}Z^{0\mu})
-2W^{+}_{\mu}W^{\mu-}A_{\nu}Z^{0\nu}\}
\end{eqnarray*}
\begin{eqnarray*}
\mathcal{L}_{ V^{2} \gamma^{2} }=e^{2}\left\{V_{\mu}^{+}V_{\nu}^{-}A^{\mu}A^{\nu}-V^{+}_{\mu}V^{\mu-}A_{\nu}A^{\nu} \right\}
\end{eqnarray*}
\begin{eqnarray*}
\mathcal{L}_{V^{2} Z^{2} }=-\frac{g^{2}}{4}\frac{(1+2s_{W}^{2})^{2}}{c_{W}^{2}}\left\{V_{\mu}^{+}V_{\nu}^{-}Z^{0\mu}Z^{0\nu}-V^{+}_{\mu}V^{\mu-}Z_{\nu}^{0}Z^{0\nu} \right\}
\end{eqnarray*}
\begin{eqnarray*}
\mathcal{L}_{V^{2} Z'^{2} }=\frac{3g^{2}}{4}\frac{(1-4s_{W}^{2})}{c_{W}^{2}}\left\{V_{\mu}^{+}V_{\nu}^{-}Z'^{\mu}Z'^{\nu}-V^{+}_{\mu}V^{\mu-}Z'_{\nu}Z'^{\nu} \right\}
\end{eqnarray*}
\begin{eqnarray*}
\mathcal{L}_{V^{2} \gamma Z }=-\frac{eg}{2c_{W}}(1+2s_{W}^{2})\{  V_{\mu}^{+}V_{\nu}^{-}(Z^{\nu}A^{\mu}+A^{\nu}Z^{\mu})\\
-2V^{+}_{\mu}V^{\mu-}A_{\nu}Z^{\nu}\}
\end{eqnarray*}
\begin{eqnarray*}
\mathcal{L}_{V^{2} \gamma Z' }=\frac{\sqrt{3}}{2}g^{2}t_{W}\sqrt{1-4s_{W}^{2}}\{  V_{\mu}^{+}V_{\nu}^{-}(Z'^{\nu}A^{\mu}+A^{\nu}Z'^{\mu})\\
-2V^{+}_{\mu}V^{\mu-}A_{\nu}Z'^{\nu}\}
\end{eqnarray*}
\begin{eqnarray*}
\mathcal{L}_{V^{2} Z Z' }=-\frac{\sqrt{3}}{4}g^{2}(1+3t_{W}^{2})\sqrt{1-4s_{W}^{2}}
\{V_{\mu}^{+}V_{\nu}^{-}(Z'^{\nu}Z^{\mu}+Z^{\nu}Z'^{\mu})\\-2V^{+}_{\mu}V^{\mu-}Z_{\nu}Z'^{\nu} \}
\end{eqnarray*}
\begin{eqnarray*}
\mathcal{L}_{U^{2} \gamma^{2} }=4e^{2}\left\{U_{\mu}^{++}U_{\nu}^{--}A^{\mu}A^{\nu}-U^{++}_{\mu}U^{\mu--}A_{\nu}A^{\nu} \right\}
\end{eqnarray*}
\begin{eqnarray*}
\mathcal{L}_{U^{2} Z^{2} }=\frac{g^{2}}{4}\frac{(1-4s_{W}^{2})^{2}}{c_{W}^{2}}
\left\{U_{\mu}^{++}U_{\nu}^{--} Z^{\mu}Z^{\nu}-U^{++}_{\mu}U^{\mu--}Z_{\nu}Z^{\nu} \right\}
\end{eqnarray*}
\begin{eqnarray*}
\mathcal{L}_{U^{2} Z'^{2} }=\frac{3g^{2}}{4}\frac{(1-4s_{W}^{2})}{c_{W}^{2}}
\left\{U_{\mu}^{++}U_{\nu}^{--}Z'^{\mu}Z'^{\nu}-U^{++}_{\mu}U^{\mu--}Z'_{\nu}Z'^{\nu} \right\}
\end{eqnarray*}
\begin{eqnarray*}
\mathcal{L}_{U^{2} \gamma Z }=\frac{eg}{c_{W}}(1-4s_{W}^{2})\{  U_{\mu}^{++}U_{\nu}^{--}(Z^{0\nu}A^{\mu}+A^{\nu}Z^{0\mu})\\
-2U^{++}_{\mu}U^{\mu--}A_{\nu}Z^{0\nu} \}
\end{eqnarray*}
\begin{eqnarray*}
\mathcal{L}_{U^{2} \gamma Z' }=\sqrt{3}g^{2}t_{W}\sqrt{1-4s_{W}^{2}}\{  U_{\mu}^{++}U_{\nu}^{--}(Z'^{\nu}A^{\mu}+A^{\nu}Z'^{\mu})\\
-2U^{++}_{\mu}U^{\mu--}A_{\nu}Z'^{\nu}\}
\end{eqnarray*}
\begin{eqnarray*}
\mathcal{L}_{U^{2} Z Z' }= \frac{\sqrt{3}}{4}g^{2}\frac{(1-4s_{W}^{2})^{3/2}}{c_{W}^{2}}\{  U_{\mu}^{++}U_{\nu}^{--}(Z'^{\nu}Z^{0\mu}+Z^{0\nu}Z'^{\mu})\\
-2U^{++}_{\mu}U^{\mu--}Z_{\nu}Z'^{\nu}\}
\end{eqnarray*}
\begin{eqnarray*}
\mathcal{L}_{WVUZ(a)}= -\frac{3\sqrt{2}}{4}\frac{g^{2}}{c_{W}}\{  W^{+}_{\mu}V^{\mu+}U^{\nu--}Z^{0}_{\nu}\\
-(1-2s_{W}^{2})W^{+}_{\mu}V^{\nu+}U^{\mu--}Z^{0}_{\nu}
-2s_{W}^{2}W^{+}_{\mu}V^{\nu+}U_{\nu}^{--}Z^{0\mu}\}
\end{eqnarray*}
\begin{eqnarray*}
\mathcal{L}_{WVUZ(b)}=-\frac{3\sqrt{2}}{4}\frac{g^{2}}{c_{W}}\{
W^{-}_{\mu}V^{\mu-}U^{\nu++}Z^{0}_{\nu}\\
-(1-2s_{W}^{2})W^{-}_{\mu}V^{\nu-}U^{\mu++}Z^{0}_{\nu}-2s_{W}^{2}W^{-}_{\mu}V^{\nu-}U_{\nu}^{++}Z^{0\mu}\}
\end{eqnarray*}
\begin{eqnarray*}
\mathcal{L}_{WVUZ'(a)}= \frac{\sqrt{6}}{4}\frac{\sqrt{1-4s_{W}^{2}}}{c_{W}}g^{2}\{  W^{+}_{\mu}V^{\mu+}U^{\nu--}Z'_{\nu}\\
+W^{+}_{\mu}V^{\nu+}U^{\mu--}Z'_{\nu}-2W^{+}_{\mu}V^{\nu+}U_{\nu}^{--}Z'^{\mu}\}
\end{eqnarray*}
\begin{eqnarray*}
\mathcal{L}_{WVUZ'(b)}= \frac{\sqrt{6}}{4}\frac{\sqrt{1-4s_{W}^{2}}}{c_{W}}g^{2}\{
W^{-}_{\mu}V^{\mu-}U^{\nu++}Z'_{\nu}\\
+W^{-}_{\mu}V^{\nu-}U^{\mu++}Z'_{\nu}-2W^{-}_{\mu}V^{\nu-}U_{\nu}^{++}Z'^{\mu}\}
\end{eqnarray*}
\begin{eqnarray*}
\mathcal{L}_{WVU\gamma (a) }=\frac{3}{\sqrt{2}}ge\left\{W_{\mu}^{+}V^{\nu+}U^{\mu--}A_{\nu}-W_{\mu}^{+}V^{\nu+}U^{--}_{\nu}A^{\mu}\right\}
\end{eqnarray*}
\begin{eqnarray*}
\mathcal{L}_{WVU\gamma (b)}=\frac{3}{\sqrt{2}}ge\left\{W_{\mu}^{-}V^{\nu-}U^{\mu++}A_{\nu}-W_{\mu}^{-}V^{\nu-}U^{++}_{\nu}A^{\mu}
\right\}
\end{eqnarray*}
\begin{eqnarray*}
\mathcal{L}_{V^{2}U^{2} }=-\frac{g^{2}}{2}\{V_{\mu}^{+}V^{\mu-}U^{++}_{\nu}U^{\nu--}+V_{\mu}^{+}V^{-}_{\nu}U^{\mu--}U^{\nu++}\\
-2V_{\mu}^{+}V^{\nu-}U^{\mu++}U^{--}_{\nu}\}
\end{eqnarray*}
\begin{eqnarray*}
\mathcal{L}_{W^{2}V^{2} }=-\frac{g^{2}}{2}\{W_{\mu}^{+}W^{\mu-}V^{+}_{\nu}V^{\nu-}+W_{\mu}^{+}W^{-}_{\nu}V^{\mu+}V^{\nu-}\\
-2W_{\mu}^{+}W^{-}_{\nu}V^{\mu-}V^{\nu+}\}
\end{eqnarray*}
\begin{eqnarray*}
\mathcal{L}_{W^{2}U^{2} }=-\frac{g^{2}}{2}\{W_{\mu}^{+}W^{\mu-}U^{++}_{\nu}U^{\nu--}+W_{\mu}^{+}W^{-}_{\nu}U^{\mu--}U^{\nu++}\\
-2W_{\mu}^{+}W^{-}_{\nu}U^{\mu++}U^{\nu--}\}
\end{eqnarray*}
\begin{eqnarray*}
\mathcal{L}_{W^{4}}=\frac{g^{2}}{2}\:W_{\mu}^{+}W_{\nu}^{-}\left\{W^{\mu+}W^{\nu-}-W^{\mu-}W^{\nu+}\right\}
\end{eqnarray*}
\begin{eqnarray*}
\mathcal{L}_{V^{4}}=\frac{g^{2}}{2}\:V_{\mu}^{+}V_{\nu}^{-}\left\{V^{\mu+}V^{\nu-}-V^{\mu-}V^{\nu+}\right\}
\end{eqnarray*}
\begin{eqnarray*}
\mathcal{L}_{U^{4}}=\frac{g^{2}}{2}\:U_{\mu}^{++}U_{\nu}^{--}\left\{U^{\mu++}U^{\nu--}-U^{\mu--}U^{\nu++}\right\}
\end{eqnarray*}

\end{document}